\documentclass[aps,prl,preprint,groupedaddress]{revtex4-1}

\def\be{\begin{equation}}
\def\ee{\end{equation}}
\def\beq{\begin{eqnarray}}
\def\eeq{\end{eqnarray}}
\def\n{\nonumber}

\usepackage{graphicx}
\begin{document}


\title{Generalised spheroidal spacetimes in 5-D Einstein--Maxwell--Gauss--Bonnet gravity}


\author{Sudan Hansraj}

\altaffiliation{}
\affiliation{Astrophysics and Cosmology Research Unit, University of KwaZulu Natal}
\email[]{hansrajs@ukzn.ac.za}


\date{\today}

\begin{abstract}

 The field equations for static EGBM gravity are obtained and transformed to an equivalent form through a coordinate redefinition. A form for one of the metric potentials that generalises the spheroidal ansatz of Vaidya--Tikekar superdense stars and additionally  prescribing the electric field intensity yields viable solutions. Some special cases of the general solution are considered and analogous classes in the Einstein framework are studied. In particular the Finch--Skea ansatz is examined in detail and found to satisfy the elementary physical requirements. These include positivity of pressure and density, the existence of a pressure free hypersurface marking the boundary, continuity with the exterior metric, a subluminal sound speed as well as the energy conditions. Moreover, the solution possesses no coordinate singularities. It is found that the impact of the Gauss--Bonnet term is to correct undesirable features in the pressure profile and sound speed index when compared to the equivalent Einstein gravity model. Furthermore graphical analyses suggest that higher densities are achievable for the same radial values when compared to the 5--dimensional Einstein case. The case of a constant gravitational potential, isothermal distribution as well as an incompressible fluid are studied. All exact  solutions derived exhibit an equation of state explicitly. 

\end{abstract}


\maketitle

\section{Introduction}

Recently exact interior metrics were reported \cite{hm, mh, chil-hans} for spherically symmetric perfect fluid distributions in the Einstein--Gauss--Bonnet (EGB) gravity theory. The exterior metric was derived by Boulware and Deser \cite{boulware} for neutral spheres and by Wiltshire \cite{wilt} for the charged counterpart a few decades ago. Since then various configurations of entities were studied in EGB gravity. The collapse of inhomogeneous dust was discussed by Jhingan and Ghosh \cite{jhingan} and a distinction was drawn between shell focussing and shell crossing hyperspheres.  Ghosh and Jhingan \cite{ghosh} examined the case of quasispherical gravitational collapse in 5D EGB framework and showed that naked singularities result and the cosmic censorship hypothesis is violated. Kleihaus {\it et al} \cite{blackstring} studied spinning black strings in the context of 5D EGB theory.    However for some three decades compact objects were not investigated in this theory on account of the highly complicated  nonlinear field equations involved. For example, the Kerr-Schild ansatz known to linearise the Einstein tensor was attempted in EGB theory \cite{anabalon} and it was found that the solution of the trace of the EGB equations did not solve all the field equations.  Success in solving the EGB equations was achieved through a coordinate transformation \cite{hm, mh, chil-hans}. This transformation was customarily used in the standard Einstein theory to convert the equation of pressure isotropy to a linear differential equation in any of the two  gravitational potentials.  While the defining master equation is not rendered as a linear differential equation, it nevertheless is possible to isolate a number of exact solutions. Examination of the resulting models revealed that they satisfy physically reasonable requirements demanded of stellar models. These include satisfying the energy conditions, the existence of a pressure free hypersurface determining the boundary of the sphere as well as the condition of causality. A similar programme proved successful in constructing compact pure Lovelock stars \cite{dad-hans}.

One of the key elements in developing a model corresponding to realistic matter is the existence of an equation of state. The equation of state for a star is known to be directly linked to the mass-radius ratio for observed objects. For example, in the work of Seager {\it {et al}} \cite{seager} mass-radius relationships for solid exoplanets were studied in an effort to understand the constituents of such planets by analysing the possible equations of state. A key finding reported in the aforesaid work is that the mass-radius relationships for cold terrestrial mass planets did not obey a simple power law form although the equations of state for solid planets are often taken to be modelled by a polytropic equation of state. In examining data on the planet HD 149026b, the mass-radius relations suggest that almost two thirds of the mass is contained within the core. \cite{sato}. The planet GJ 436b displayed mass-radius behaviour akin to Neptune \cite{butler, gillon}. Commencing with a prescribed equation of state imposes severe restrictions on the system of nonlinear differential equations and impedes the finding of exact solutions \cite{nils1,nils2}. What has been found is that equations of state may more readily be found when other mathematical prescriptions are made to complete the model. For example see the Finch--Skea \cite{fs} model (and its charged counterpart \cite{hansraj})  which has been shown to be compatible with astrophysical predictions in the theory of Walecka \cite{wal}. In the current problem, the field equations are more formidable yet we find that the equation of state may be determined. In fact, this strongly suggests that in constructing exact models of stars, the route of specifying a geometrical component and/ or another aspects of the physics (such as the electric field in our case), is a more productive means of determining the equation of state.

Charged fluid distributions have been traditionally studied despite the notion that celestial bodies are generally understood to be neutral. In earlier times efforts to apply solutions of the Einstein--Maxwell system to model the electron failed on account of an unrealistic mass-radius ratios. Nevertheless, it is mathematically worthwhile to examine the possibilities for charged compact objects to exist in modified theories of gravity. We attempt to construct such models in the present work. Since the problem in EMGB is similar to the Einstein theory in that there are 4 equations governing six unknowns which means that effectively there are 15 different two element sets of these six variables that may be chosen a priori to close the system.   A listing of the various two element choices that have been made historically is contained in \cite{ivan}. As in the simplified Einstein case, it is a trivial exercise to find exact solutions if the two metric potentials are prescribed at the outset. However, it is by no means trivial if other two choices are made besides both metric potentials. For example, the following problems are under study. Models of charged dust by setting the pressure to zero then require one further prescription and the process is by no means trivial. Such a model is interesting as Coulombic forces oppose the pressure to prevent the collapse of the star to a point singularity. Then if a linear barotropic equation of state is imposed, one further choice needs to be made to close the system. Again, this is a nontrivial problem and will be discussed elsewhere.

The EGB action principle is a quadratic modification of the Einstein action which is linear in the Ricci tensor. Here quadratic forms of the Riemann tensor, Ricci tensor and Ricci scalar are used to construct an action principle that yields at most second order equations of motion. EGB belongs to a larger class of $N^{th}$ order polynomials discovered by Lovelock \cite{lovelock} that represent the most general Lagrangians that generate up to second order differential equations governing the evolution of stars. EGB is the $N=2$ case of the Lovelock polynomial. Vacuum solutions have been determined for Lovelock gravity in general \cite{whit, whee} as well as for dimensionally continued \cite{Ban1,Ban2,Myers,wilt}, and also for pure Lovelock black holes \cite{probes3,cai}.   Sharma \cite{Sharm}  showed that the limiting case of the polytropic fluid may be regarded as isothermal and is important in the study of clusters of stars.
 It is demonstrated by Dadhich and coworkers \cite{dkm} that the Schwarzschild interior solution is universal in describing a uniform density sphere in all dimensions $>3$ in higher dimensional Einstein or Lovelock theory.
In fact Dadhich \cite{dad} has strongly argued that the pure Lovelock theory in which only the $N^{th}$ order term (and not the sum of terms  up to $N$) constitute the more accurate theory of gravity. Indeed the $N=1$ case is general relativity so the   results in general relativity still hold for that special case.

 Higher derivative gravity theories have been proposed in order to explain the cosmic accelerated expansion without resorting to exotic matter. Strong support for EGB theory lies in the fact that the effective action in heterotic string theory involves a Gauss-Bonnet term \cite{gross}. Of course string theory relies upon the existence of higher dimensions than 4 - something which is generally unpalatable in a theory of gravitation. However, if a grand unified theory exists then higher dimensional effects must exist in gravity theory as they do in the quantum regime - whether they are accessible is an open question. Generally, higher dimensions are explained away by claiming that they are hidden topologically. These issues were also considered in \cite{dad}.

 There are other theories of the gravitational field gaining popularity.The  trace-free Einstein equations (also known as unimodular gravity) have been proposed by Ellis \cite{ellis1,ellis2} as the true theory of gravity. This idea, first proposed by Weinberg \cite{weinberg}, meant to address the  great difference in the value of the vacuum energy density predicted by quantum theory in comparison  to the actual value of the cosmological constant from astronomical observations. Starobinsky's $f(R)$ theory \cite{staro} is also well studied, however, unlike Lovelock theory fourth order equations of motion emerge. The theory has been shown to be conformally equivalent to scalar tensor theory plus general relativity. Of late $f(R, T)$ theory \cite{harko} has been proposed where the Lagrangian is functionally dependent on the Ricci scalar as well as the trace of the energy momentum tensor $T_{ab}$. These investigations show that modified theories of gravity are indeed worthwhile areas of study and corrections to the standard theory may result. It is also important to note that studies of data collected by observations of celestial phenomena are grounded in the results of general relativity. If such computer codes were written using the modifications to the theory (such as the EGB corrections herein) the results that emerge may well be different and could convey a different portrait of the universe, stellar structures and galaxy formation and evolution. In this spirit, it is believed that the implications of extended theories are important.

 In this paper we develop the governing equations for a spherically symmetric distribution of matter coupled with an electric field in the  EGB framework. Exact solutions for the resulting Einstein--Maxwell-Gauss--Bonnet (EMGB) field  equations are then sought. The equation of pressure isotropy is recast into an equivalent form with the help of a coordinate transformation. The analysis amounts to solving a system of four partial differential equations in six unknowns: energy density $\rho$, pressure $p$, electric field intensity $E$, proper charge density $\sigma$ and two metric potentials $\nu$ and $\lambda$. As is characteristic of the EGB framework the isotropy equation is second order in one variable and first order in the other. Unlike the Einstein theory, nonlinearity is inherent in both formulations of the isotropy equation.

Ordinarily an equation of state is selected to help close the system however one further assumption may be made to generate a unique model. This has generally proved a difficult route in the standard Einstein gravity. An alternative viable route is to propose functional forms for two of the six variables and endeavor to solve the system. Note that the four dynamical variables may all be expressed in terms of the metric potentials $\nu$ and $\lambda$ and so specifying the geometry will trivially produce a unique model without needing to perform any integrations. This approach has been followed by  \cite{krori} to generate a singularity free model  but the drawback is that all control over the pressure and density is sacrificed. We elect to specify one of the metric potentials and the electric field intensity in order to solve the system of field equations. This approach has been shown to yield physically viable charged star models in Einstein theory and importantly consisting of a barotropic equation of state \cite{hansraj}. Therefore this route is being pursued in this work.  The form of the potential $\lambda$ selected corresponds to a generalised version of a spheroidal spacetime and contains a number of well studied ansatze in the Einstein theory and we investigate their behavior in this modified EMGB theory.

The paper is arranged as follows: We recall basic elements of EGB gravity and then extend the theory to incorporate the electric field by supplementing the field equations with the Maxwell's equations. The field equations for a static spherically symmetric perfect fluid in a charged field are written and converted to an equivalent form. We then postulate a general form for the metric potential $\lambda$ and the electric field that permits the complete integration of the field equations. Special cases of the potential are then considered and the exact model is exhibited in these cases.


\section{Einstein--Gauss--Bonnet Gravity}

The action principle in the standard Einstein theory of relativity  is the Einstein--Hilbert  action given by
\begin{equation}
S=\frac{1}{2\kappa} \int R \sqrt{-g} d^4 x \label{1}
\end{equation}
where $g = \det (g_{ab}) $ is the determinant of the metric tensor $g_{ab}$, $R$ is the Ricci scalar and $\kappa = 8\pi Gc^{-4}$ where $G$ is the Newton's gravitational constant and $c$ is the speed of light in vacuum. The Einstein field equations
 \begin{equation}
 R_{ab} - \frac{1}{2} g_{ab} R = \kappa T_{ab} \label{2}
 \end{equation}
consequently arise where $T_{ab} $ is the energy momentum tensor for the matter configuration. If the   cosmological constant $\Lambda$ is included then the Lagrangian has the form
\begin{equation}
S= \int \frac{1}{2\kappa}\left( R - 2\Lambda\right)  \sqrt{-g} d^4 x \label{3}
\end{equation}
and the associated Einstein field equations are given by
\begin{equation}
R_{ab} - \frac{1}{2} g_{ab} R + \Lambda g_{ab} = \kappa T_{ab}. \label{4}
\end{equation}

The Lovelock \cite{lovelock} Lagrangian is written as
\begin{equation}
\mathcal{L} = \sum ^t_{n=0} \alpha_n \mathcal{R}^n  \label{5}
\end{equation}
where $ \mathcal{R}^n = \frac{1}{2^n} \delta^{c_1 d_1 ...c_n d_n}_{a_1 b_1 ... a_n b_n} \Pi^n_{r=1} R^{a_r b_r}_{c_r d_r} $ and $R^{a b}_{c d}$ is the Riemann or curvature tensor. Also $ \delta^{c_1 d_1 ...c_n d_n}_{a_1 b_1 ... a_n b_n} = \frac{1}{n!} \delta^{c_1}_{\left[a_1\right.} \delta^{d_1}_{b_1} ... \delta^{c_n}_{a_n} \delta^{d_n}_{\left.b_n \right]}$ is the required Kronecker delta.

The Lovelock action (\ref{5}) may be expanded as
\[
\mathcal{L} = \sqrt{-g}\left( \alpha_0 + \alpha_1R + \alpha_2 \left( R^2 + R_{a b c d} R^{a b c d} - 4R_{c d} R^{c d} \right) + \alpha_3 \mathcal{O}(R^3)\right)
\]
from which we define the Gauss--Bonnet (GB) term as
\[
\mathcal{R}^2 = R^2 + R_{a b c d} R^{a b c d} - 4R_{c d} R^{c d}
\]
denoted as $L{GB}$.
This term  arises in the low energy effective action of heterotic string theory \cite{gross}.
As a result the Einstein--Gauss--Bonnet field equations are given by
\begin{equation}
G^{a}_{b} + \alpha H^{a}_{b} = T^{a}_{b}  \label{6}
\end{equation}
where
\[
H_{ab} = 2\left(R R_{ab} - 2R_{ac}R^c_b - 2R^{cd}R_{acbd} + R^{cde}_{a} R_{bcde} \right) - \frac{1}{2} g_{ab} \mathcal{R}^2.
\]
The Gauss--Bonnet action is written as
\begin{equation}
S = \int \sqrt{-g} \left[ \frac{1}{2} \left(R - 2\Lambda + \alpha L_{GB}\right)\right] d^n x + S_{\mbox{ matter}} \label{7}
\end{equation}
where $\alpha $ is the GB coupling constant. The constant $\alpha$ is linked with the string tension in string theory \cite{gross}. The remarkable feature  of the GB action lies in the fact that despite the Lagrangian being quadratic in the Ricci tensor, Ricci Scalar  and the Riemann tensor, the equations of motion turn out to be second order quasilinear.  The GB term has no effect for $n \leq 4$ but is generally non--zero for $n > 4$.

\section{Field equations for charged spheres}

The generic 5-D  line element for static spherically symmetric spacetimes may be expressed as
\begin{equation}
ds^{2} = -e^{2 \nu} dt^{2} + e^{2 \lambda} dr^{2} + r^{2} \left( d\theta^{2} + \sin^{2} \theta d \phi^2 + \sin^{2} \theta \sin^{2} \phi d\psi^2 \right) \label{8}
\end{equation}
where $ \nu(r) $ and $ \lambda(r) $ are  the gravitational potentials.
                     The Einstein--Maxwell-Gauss-Bonnet (EMGB)  system of field equations  is given by
\beq
G_{ab} &=& T_{ab}               \n \\ \n \\
       &=& M_{ab} + E_{ab}       \label{511a} \\ \n \\
F_{ab;c} + F_{bc;a} + F_{ca;b} &=& 0    \label{511b} \\ \n \\
F^{ab}{}{}_{;b} &=& J^a \label{511c}
\eeq
where   ${\bf T}$ is the total  energy--momentum tensor,
${\bf M}$ is the energy--momentum tensor for neutral matter, ${\bf E}$ is the
contribution of the electromagnetic field,
${\bf F}$ is the electromagnetic field
tensor  and
 ${\bf J}$
is the five--current density.

The electromagnetic
contribution {\bf E} to the total energy--momentum tensor is given by
\be
E_{ab} = F_{ac}F_b{}^{c} - \frac{1}{4}g_{ab}F_{cd}F^{cd} \label{514}
\ee
where ${\bf F}$ is  skew--symmetric.
The five--current density for a non--conducting fluid  can be written as
\be
J^a=\sigma u^a    \label{514'}
\ee
where $\sigma$
is the proper charge density.
The electromagnetic field tensor ${\bf F}$  is defined in terms of the
five--potential ${\bf A}$ by
\be
F_{ab} = A_{b;a} - A_{a;b} \label{515}
\ee
Gauge freedom allows us to
choose the five--potential as \[
          A_a = (\phi(r),0,0,0,0)   \nonumber
\]
so that the effects of the magnetic field are suppressed and only the electric field contributes.
Only  one non-zero component of the Faraday tensor $F_{ab}$ survives and is given by
\be
          F_{01} = -\phi'(r)  \label{516}
\ee
where we have utilised (\ref{515}).
The corresponding contravariant component
has the form
\[
         F^{01} = e^{-2(\nu+\lambda)}\phi'(r)
                = e^{-(\nu+\lambda)}E(r)          \nonumber
\]
where we have put
\be
E(r) = e^{-(\nu+\lambda)}\phi'(r)   \label{517}
\ee
following  Herrera and Ponce de Leon \cite{herrera} for the 4 dimensional Einstein--Maxwell equations. The quantity
$E(r)$ is  interpreted as the electrostatic field intensity.
 The components of the
 electromagnetic field energy tensor (\ref{514})       are  given by
\be
E^a_{b}=\mbox{diag} \left( -\frac{1}{2}E^2,\,
-\frac{1}{2}E^2,\,
\frac{1}{2}E^2,\,  \frac{1}{2}E^2, \frac{1}{2} E^2 \right)
\label{523} \ee
where we have used (\ref{516}).
 The energy--momentum tensor ${\bf M}$ for the
comoving fluid
velocity vector $u^a=e^{-\nu} \delta^{a}_{0}$
has the form
\be
M^a_{b}= \mbox{diag} \left( -\rho,\, p ,\,  p ,\,
p,\, p\right) \label{518}
\ee
for uncharged matter.

\vspace{5mm}
The field equation (\ref{511c}) gives the relationship
 \be
e^{-\lambda}\left(r^2 E\right)'=r^2 \sigma \label{522}
\ee
for the value $a = 0$.
The conservation laws
  $T^{ab}_{}{}_{;b} = 0 $
generate the equation
\beq
p' + (\rho + p)\nu' =
               \frac{E}{r^2}\left[r^2 E\right]' \label{525}
\eeq
which can be substituted for  one of the field equations.

 For this line element the EMGB field equations are given by
\begin{eqnarray}
\rho + \frac{1}{2}E^2 &=& -\frac{3}{e^{4 \lambda} r^{3}} \left( 4 \alpha \lambda ^{'} +  r e^{2 \lambda} -  r e^{4 \lambda} -  r^{2} e^{2 \lambda} \lambda ^{'} - 4 \alpha e^{2 \lambda} \lambda ^{'} \right)  \label{9a} \\ \nonumber \\
p -\frac{1}{2} E^2  &=&  \frac{3}{e^{4 \lambda} r^{3}} \left(-  r e^{4 \lambda} + \left( r^{2} \nu^{'} +  r + 4 \alpha \nu^{'} \right) e^{2 \lambda} - 3 \alpha \nu^{'} \right) \label{9b} \\ \nonumber \\
p + \frac{1}{2} E^2  &=& \frac{1}{e^{4 \lambda} r^{2}} \left( -e^{4 \lambda} - 4 \alpha \nu^{''} + 12 \alpha \nu^{'} \lambda^{'} - 4 \alpha \left( \nu^{'} \right)^{2}  \right) \nonumber \\
                 & \quad & + \frac{1}{e^{2 \lambda} r^{2}} \left(  1 - r^{2} \nu^{'} \lambda^{'} + 2 r \nu^{'} - 2 r \lambda^{'} + r^{2} \left( \nu^{'} \right)^{2}  \right) \nonumber \\
                 & \quad & + \frac{1}{e^{2 \lambda} r^{2}} \left(  r^{2} \nu^{''} - 4 \alpha \nu^{'} \lambda^{'} + 4 \alpha \left( \nu^{'} \right) ^{2} + 4 \alpha \nu^{''}   \right). \label{9c} \\ \nonumber \\
                  e^{-\lambda}\left(r^2 E\right)' &=& r^2 \sigma \label{9d}
\end{eqnarray}
in the canonical spherical coordinates.

Employing  the transformations  $ e^{2 \nu} = y^{2}(x) $, $ e^{-2 \lambda} = Z(x)  $ and $ x = C r^{2}$ ($C$ a constant) the field equations (\ref{9a} -- \ref{9d}) may be rewritten as
\begin{eqnarray}
 \frac{3  (1-Z) ( 1 - 4 \alpha C \dot{Z} )}{x} -3\dot{Z}  &=& \frac{\rho}{C} + \frac{E^2}{2C} \label{10a} \\ \nonumber \\
\frac{3  (Z - 1)}{x} + \frac{6  Z \dot{y}}{y} - \frac{24 \alpha C (Z - 1) Z \dot{y}}{x y} &=& \frac{p}{C} - \frac{E^2}{2C} \label{10b} \\ \nonumber \\
 2 x Z \left( 4 \alpha C [1-Z] + x \right) \ddot{y}  + \left( x^{2} \dot{Z} + 4 \alpha C \left[ x \dot{Z} - 2 Z + 2 Z^{2} - 3 x Z \dot{Z} \right] \right) \dot{y} \nonumber \\ + \left( 1 + x \dot{Z} - Z -\frac{E^2}{C} x \right) y &=& 0 \label{10c}\\ \nonumber \\
\frac{4Z}{x}\left( x\dot{E} + E \right)^2 &=& \frac{\sigma^2}{C}
\label{10d}
\end{eqnarray}
where (\ref{10c}) is the equation of pressure isotropy. Equation (\ref{10c}) has been arranged as a second order differential equation in $y$.  However, it should be noted that (\ref{10c}) may also be regarded as a first order ordinary differential equation in $Z$. Essentially the system (\ref{10a}) -- (\ref{10d}) comprise an under-determined system of four coupled partial differential equations in six unknowns. To complete the system two choices for the geometric or matter variables may be made or alternatively a functional dependence of one quantity on another, such as an equation of state, may be selected and then a second choice made. Clearly all the matter variables may be written explicitly in terms of the metric functions $Z$ and $y$. This means that any metric can satisfy the EMGB field equations. This is also true for the simpler Einstein--Maxwell field equations. However, arbitrary choices of metrics will not easily yield dynamical variables that may be considered physically reasonable. It would hardly be expected that an equation of state may exist if random metrics are selected. Prescribing an equation of state then creates the problem of an intractable system - nevertheless this is a viable direction that is being pursued by the author in a different work. For the purposes of this investigation we elect to nominate an electric field intensity that behaves essentially as the reciprocal of the radius to conform with the Newtonian case. The metric ansatz chosen, namely the Vaiyda--Tikekar one, has been extensively studied and shown to generate models that do satisfy the elementary requirements for physical acceptability. Hence, this is a route we pursue in order to study the effect of the higher curvature terms on the gravitational behaviour of the charged fluid when compared to general relativity (GR).

Note  that on setting $\alpha = 0$ in the system (\ref{10a}) -- (\ref{10d}) these equations take the form
\begin{eqnarray}
 \frac{3  (1-Z) }{x} -3\dot{Z}  &=& \frac{\rho}{C} + \frac{E^2}{2C} \label{10a1} \\ \nonumber \\
\frac{3  (Z - 1)}{x} + \frac{6  Z \dot{y}}{y}  &=& \frac{p}{C} - \frac{E^2}{2C} \label{10b1} \\ \nonumber \\
 2 x^2 Z  \ddot{y}  +  x^{2} \dot{Z}  \dot{y} +\left( \dot{Z} x - Z +  1 - \frac{E^2}{C} x \right) y &=& 0 \label{10c1}\\ \nonumber \\
\frac{4Z}{x}\left( x\dot{E} + E \right)^2 &=& \frac{\sigma^2}{C}
\label{10d1}
\end{eqnarray}
which constitute the 5-dimensional Einstein equations for static charged perfect fluid spheres.  In what follows we report new exact models for the EMGB field equations and then analyse the role of the Gauss--Bonnet coupling by setting it to zero to facilitate a comparison with the 5--D Einstein charged spacetime.

The following conditions are usually imposed on models as elementary physical requirements. It is required that the energy density ($\rho$) and  pressure ($p$) are positive. The pressure should vanish for some radial value $r = R$. The sound speed should be less than the speed of light, that is, $0 < \frac{dp}{d\rho} < 1$. Across the boundary $r=R$ the interior metric should match with the exterior charged Boulware--Deser  spacetime \cite{wilt}
\be
ds^2 = -{\cal{F}}(r) dt^2 + \frac{dr^2}{{\cal{F}}} + r^{2} \left( d\theta^{2} + \sin^{2} \theta d \phi^2 + \sin^{2} \theta \sin^{2} \phi d\psi^2 \right) \label{008}
\ee
where ${\cal{F}}(r) = K + \frac{r^2}{4\alpha} \left(1-\sqrt{1 + \frac{8\alpha M}{r^4} - \frac{8\alpha Q^2}{3r^6}} \right)$
in the absence of the cosmological constant. Recall that $\alpha$ is a coupling constant related to the string tension in string theory and $K$ is an arbitrary constant.  Note that $M$ and $Q$ represent the gravitational mass and charge of the fluid as measured by an observer at spatial infinity. It should be disclosed that these boundary conditions are extrapolated from the standard Einstein case. The junction conditions for the EGB framework was discussed by Davis \cite{davis} however these results have yet to be transformed to an explicit useable form for modelling purposes.

\section{Generalised Vaidya--Tikekar ansatz}

As the EMGB field equations constitute a system of four  field equations in six unknowns, functional forms for any two of the matter or geometrical quantities may be postulated and by integration the remaining variables may be determined. The alternative approach is to specify an equation of state relating the isotropic particle pressure and the energy density, that is $p=p(\rho)$. This approach has has given rise to difficulties in the simpler version of the Einstein's equations therefore is not pursued in  this framework at present. For example, see the work of Nillson and Uggla \cite{nils1,nils2} where the analysis of both the barotropic gamma law equation of state as well as the polytropic distribution was discussed numerically in the context of general relativity. Exact solutions were not located.  A more productive approach appears to be selecting a form of one of the gravitational potentials $y$ or $Z$ and also prescribing how the electric field $E$ behaves. This is the route followed here.

In our investigation we study the gravitational potential in the form
\be
e^{-2\lambda} = Z=\frac{1+ax}{1+bx} \label{001}
\ee
where $a$ and $b$ are two real parameters. This may be recognised as a generalisation of the Vaidya--Tikekar ansatz \cite{vt}. Setting $b=1$ then the parameter $a$ is understood as the spheroidal parameter as discussed in \cite{vt}and which has been used in the construction of models of superdense stars. Setting $a=0$ and $b=1$  regains the Finch--Skea metric \cite{fs}. The case $b=0$ and $a=1$ corresponds to the Schwarzschild interior metric in Einstein gravity and which has been shown to be a persistent solution in other theories of gravity such as Lovelock gravity \cite{lovelock}. This case deserves special consideration and will be considered in a different article.  Putting $a=b$ reduces the metric to a Minkowski spacetime for $Z=1$. More general constant gravitational potentials deserve detailed treatment in their own right as they may generate isothermal fluid spheres with the Newtonian behaviour density falling off according to the inverse square law and a linear barotropic equation of state. It has been shown that isothermal fluids are universal in Lovelock gravity \cite{isothermal} and that the necessary and sufficient condition for isothermal behaviour is a constant gravitational potential. The Einstein version \cite{sas} was regained as a special case.

 With the general ansatz (\ref{001}), the master field equation (\ref{10c}) assumes the form
\beq
2x(1+ax)(1+bx)(1-ak+b(k+x))\ddot{y} + (a-b)x(1+3bk+bx+ak(2bx-1))\dot{y} \nonumber \\
-(1+bx)\left((a-b)bx +(1+bx)^2\frac{E^2}{C}\right)y=0 \label{002}
\eeq
where we have put $4\alpha C = k$. It now remains to select suitable forms for the electrostatic function $E$ to allow for the complete integration of (\ref{002}).
The choice
\be
\frac{E^2}{C} = \frac{(b-a)bx}{(1+bx)^2} \label{003}
\ee
for the electric field is motivated by the fact that the last term on the left of (\ref{002}) vanishes thus effectively reducing the order of the differential equation to first order. Moreover it is a physically viable choice and has the property that the electric field disappears at the stellar centre $x=0$. Also the right hand side is positive provided $b(a-b)<0$. The electric field has a maximum or minimum value (depending on whether $a > b$ or not, of $\frac{C(b-a)}{4}$ when $x=\frac{1}{b}$.

With this prescription, the field equation (\ref{002}) is solved exactly by
\beq
y(x) &=& c_1 \left(\frac{(a-b) (2 a k+1) \log \left(2 a b x+2 \sqrt{ab(1+ax)(1+bx)}+a+b\right)}{2 a^{3/2} \sqrt{b}} \right. \\ \nonumber \\ \nonumber &&\left.
+\frac{k^{3/2} (a-b) \log (-a k+b (k+x)+1)}{\sqrt{b(ak-1)}} \right. \nonumber \\ \nonumber \\ && \left.
-\frac{k^{3/2} (a-b) \log \left(\sqrt{b} \left(2 a b k x+2 \sqrt{bk(ak-1)(1+ax)(1+bx)} + (a+b) k -b x-1\right)\right)}{\sqrt{b(ak-1)}} \right. \nonumber \\ \nonumber \\ && \left.
+\frac{\sqrt{(1+ax)(1+bx)} }{a}\right)+c_2 \label{004}
\eeq
where $c_1$ and $c_2$ are constants of integration. Note immediately that the choices $a=0$ and $b=0$ are excluded from the solution (\ref{004}). These cases must be treated separately. The energy density, pressure and proper charge density may now be obtained with the help of (\ref{10a}), (\ref{10b}) and (\ref{10d}), however, given the lengthy form of the solution (\ref{004}), it will be instructive to examine the physical behaviour for specific cases with a known Einstein analogue.

\subsection{Vaidya--Tikekar case $a=1$, $b=2$}

The case $Z=\frac{1+2x}{1+x}$ was investigated thoroughly by Vaidya and Tikekar \cite{vt} for a four dimensional perfect fluid source without charge. The electric field in our case simplifies to
\be
\frac{E^2}{C} = \frac{2x}{(1+2x)^2} \label{006}
\ee
while the gravitational potential has the form
\beq
y &=& c_1 \left(\frac{k^{3/2}}{ \sqrt{k-1}} \log \frac{ \left(\sqrt{2} \left(k(4  x+ 3) + 2 \sqrt{2(k-1)k(x+1)(2 x+1)} -(2 x+1)\right)\right)}{ (2 (k+x)-k+1)} \right.  \nonumber \\ && \left.
-\frac{(2 k+1) \log \left(4 x+2 \sqrt{2(x+1)(2 x+1)}+3\right)}{2 }+\sqrt{2(x+1)(2 x+1)}\right)+c_2  \label{0006b}
\eeq
for $k\neq 1$. The energy density is given by
\be
\frac{\rho}{C} =\frac{ \left(3 k+10 x^2+17 x+6\right)}{(2 x+1)^3} \label{006c}
\ee
while the pressure has the form

\be
\frac{p}{C} = \frac{x}{(2 x+1)^2}-\frac{3}{2 x+1} -\frac{24 c_1 \sqrt{x+1} \, V}{ \sqrt{2x+1} \, L H   S } \label{006d}
\ee
where we have made the redefinitions \\
$H=\left(k (4 x+3)   -(2 x+1) +2   \sqrt{2k(k-1)(x+1)(2x+1)}   \right)$ \\
$J=\left(8 x^2+4 \left( \sqrt{2(x+1)(2x+1)} +3\right) x  +3  \sqrt{2(x+1)(2x+1)}+4\right)$ \\
$L=\left(4 x+ 3 + 2  \sqrt{2(x+1)(2x+1)} \right)$ \\
$S= 2 \sqrt{2} c_1 k^{3/2} \log \frac{(k+2 x+1)}{\sqrt{2}H}    - 2 \sqrt{k-1} \left(c_1 \sqrt{(x+1)(2x+1)} +c_2\right)
 - \sqrt{2} c_1 (2 k+1) \sqrt{k-1} \log L$ \\
 $V = 2 \sqrt{k} J(k-1) +\sqrt{k-1}L \left( k (4 x+3)   -  (2 x+1) \right) $
for simplicity. The expressions for the sound speed is easily achievable but is omitted for brevity. Similarly it is straightforward to write the constants $c_1$ and $c_2$ in terms of $M$, $Q$ and $R$ using the zero boundary pressure  condition $p(R)=0$ as well as the continuity of $g_{11}$ across $r=R$ since this amounts to solving two algebraic equations linear in $c_1$ and $c_2$  however we do not display the lengthy expressions. An important observation in this model is that an equation of state exists and is easily calculated. From equation (\ref{006c}) $x$ may be expressed in terms of $\rho$ and the resulting expression for $\rho$ may be substituted in (\ref{006d}) to give the functional dependence of pressure on density as desired.

\subsection{Finch--Skea case}

The Finch--Skea \cite{fs} case is obtained for $a=0$ and $b=1$ that is $Z = \frac{1}{1+x}$. In this case the other gravitational potential is given by
\be
y=\frac{2}{3} c_1 \left(3 k^{3/2} \tan ^{-1}\left(\frac{\sqrt{x+1}}{\sqrt{k}}\right)+\sqrt{x+1} (-3 k+x+1)\right)+c_2 \label{007a}
\ee
for integration constants $c_1$ and $c_2$. This case is of particular importance as it has been demonstrated to correspond to realistic stellar distributions in Einstein gravity according to the theory of Walecka \cite{wal}.
The electric field intensity is given by
\be
\frac{E^2}{C} =\frac{x}{(x+1)^2} =\frac{Cr^2}{(1+Cr^2)^2} \label{007b}
\ee
and correspondingly the proper charge density assumes the form
\be
\frac{\sigma^2}{C} = \frac{C (x+3)^2}{(x+1)^5} = \frac{C(3 +Cr^2)^2}{(1+Cr^2)^5}\label{007b1}
\ee
The energy density and pressure have the forms
\beq
\frac{\rho}{C} &=& \frac{6 k+5 x^2+17 x+12}{2 (x+1)^3} \label{007c} \\ \n \\
\frac{p}{C} &=& \frac{2 c_1 (x+1) \left(3 k (5 x+6)-5 x^2+7 x+12\right) -3 c_2 \sqrt{x+1} (5 x+6)}{2 (x+1)^{5/2} \left(6 c_1 k^{3/2} \tan ^{-1}\left(\frac{\sqrt{x+1}}{\sqrt{k}}\right)+2 c_1 \sqrt{x+1} (-3 k+x+1)+3 c_2\right)} \nonumber  \\ && - \frac{ 6 c_1 k^{3/2} \sqrt{x+1} (5 x+6) \tan ^{-1}\left(\frac{\sqrt{x+1}}{\sqrt{k}}\right)}{2 (x+1)^{5/2} \left(6 c_1 k^{3/2} \tan ^{-1}\left(\frac{\sqrt{x+1}}{\sqrt{k}}\right)+2 c_1 \sqrt{x+1} (-3 k+x+1)+3 c_2\right)} \label{007d}
\eeq
respectively. Note that solving the cubic equation (\ref{007c} for $x$ and plugging into (\ref{007d}) gives a barotropic equation of state for this model. The equation of state is explicitly given by 
\beq
p &=& 18\sqrt{6}\left(60-3\sqrt{6}h -\frac{5}{108\rho^3} \left(h^2 -6\rho\right)^3 + \frac{17}{18\rho^2} \left( 6\rho - h^2\right)^2 +\frac{1}{6\rho} \left(h^2 - 6\rho\right) \left(104 - 5\sqrt{\frac{3}{2\rho^2}} h\right) \right. \n \\ \n \\
&& \left. -\sqrt{\frac{6}{\rho^2}} h \left( 6 + \frac{5}{6\rho} \left(h^2 -6\rho\right)\right)
\tanh^{-1} \frac{h}{\sqrt{6\rho}}  \right) \n \\ \n \\
&&
/ \left(\left(\frac{1}{\rho} h^2\right)^{\frac{5}{2}} \left(3 -\sqrt{\frac{8}{3\rho^2}} h + \frac{1}{3\sqrt{6}\rho^3} h  \left( h^2 -6\rho  \right) + 6\tanh^{-1} \frac{h}{\sqrt{6\rho}}               \right) \right) \label{007b2}
  \eeq
where we have substituted 
\beq
h &=& \left(-\sqrt[3]{3 \sqrt{3} \sqrt{\rho ^2 (16 \rho  (243 \rho +301)+1775)}-9 \rho  (36 \rho
+35)-125} \right. \n \\ \n  \\
&& \left. +\frac{-42 \rho -25}{\sqrt[3]{3 \sqrt{3} \sqrt{\rho ^2 (16 \rho  (243 \rho +301)+1775)}-9 \rho  (36 \rho +35)-125}}+5\right)^{1/2}
\eeq
to shorten the lengthy expression for $p(\rho)$. Although cumbersome and complicated the equation of state does indeed exist and this underscores the methodology used that of specifying a gravitational potential and the electric field intensity. If an equation of state were imposed early in the model the mathematical complexity may not have permitted a complete discovery of the model. 

The sound speed may be expressed as
\beq
\frac{dp}{d\rho} &=& -\left\{(x+1) \left(12 c_1 k^{3/2} (k+x+1) \left(c_1 \sqrt{x+1} \left(5 \left(2 x^2+3 x+1\right)-6 k (5 x+7)\right) \right. \right. \right.  \n \\ \n \\
&&  \left. \left. \left. +3 c_2 (5 x+7)\right) \tan ^{-1}\left(\frac{\sqrt{x+1}}{\sqrt{k}}\right)+36 c_1^2 k^3 (5 x+7) (k+x+1) \tan ^{-1}\left(\frac{\sqrt{x+1}}{\sqrt{k}}\right)^2 \right. \right.  \n \\ \n \\
&&  \left. \left.  +6 c_1 c_2 \sqrt{x+1} \left(-6 k^2 (5 x+7)-k \left(20 x^2+57 x+37\right)+5 (x+1)^2 (2 x+1)\right)    \right. \right.  \n \\ \n \\
&& \left. \left. +4 c_1^2 (x+1) \left(9 k^3 (5 x+7)+3 k^2 \left(5 x^2+21 x+16\right)-k (x+1)^2 (25 x+17 ) \right. \right. \right.  \n \\ \n \\
&& \left. \left. \left.
 +(x+1)^3 (5 x-29)\right)+9 c_2^2 (5 x+7) (k+x+1)\right)\right\}/ \left\{(k+x+1) \left(18 k+5 x^2+24 x+19\right) \right.  \n \\ \n \\
 && \left. \left(6 c_1 k^{3/2} \tan ^{-1}\left(\frac{\sqrt{x+1}}{\sqrt{k}}\right)+2 c_1 \sqrt{x+1} (-3 k+x+1)+3 c_2\right){}^2\right\}  \label{007e}
\eeq
The mass function is computed from $m(r)=\int \rho(r) r^2 dr$ and is given by
\beq
(m(r))_{EGB} &=& \frac{1}{8} \left(\frac{r \left(20 r^4+29 r^2+3\right)}{\left(r^2+1\right)^2}-3 \tan ^{-1}(r)\right)   \label{007e1} \\ \n \\
(m(r))_{GR} &=& \frac{1}{2} \left(-\frac{7 r}{2 \left(r^2+1\right)}+5 r-\frac{3}{2} \tan ^{-1}(r)\right) \label{007e2}
\eeq
for the EGB and GR frameworks respectively.
Solving the vanishing boundary condition and the continuity of metric potentials across $r=R$ we obtain
\beq
c_1 &=& \frac{(5X+1) \left(-3 k K+(X-1)\left(\sqrt{3}  \sqrt{\frac{24 \alpha  M R^2-8 \alpha  Q^2+3 R^6}{R^6}}-3 \right)\right)}{k \sqrt{X} \left(f-36X\right)} \\ \n \\
c_2 &=&  \frac{\left(30 k^{3/2} (X-1) \tan ^{-1}\left(\frac{\sqrt{X}}{\sqrt{k}}\right)X -f\sqrt{X}\right) \left(3 k K- (X-1)\left(\sqrt{3}  \sqrt{\frac{24 \alpha  M R^2-8 \alpha  Q^2+3 R^6}{R^6}}+3 \right)\right)}{3 k \sqrt{X} \left(f-36X\right)} \n \\
\label{007g}
\eeq
for the integration constants and where we have put $X=1+CR^2$ and \\ $f=15 k (X-1) +18 k -5 X^2 +17X $. Note that a boundary certainly exists in this model as setting $p=0$ amounts to solving a cubic algebraic equation and the existence of at least one real-valued solution is guaranteed.
In order to examine the physical properties of our model graphically we select the following parameter values: $c_1 =c_2= k=C=1$. Plots of the dynamical quantities have been generated with the help of Mathematica 11.

\begin{figure}
  \includegraphics[width=8cm]{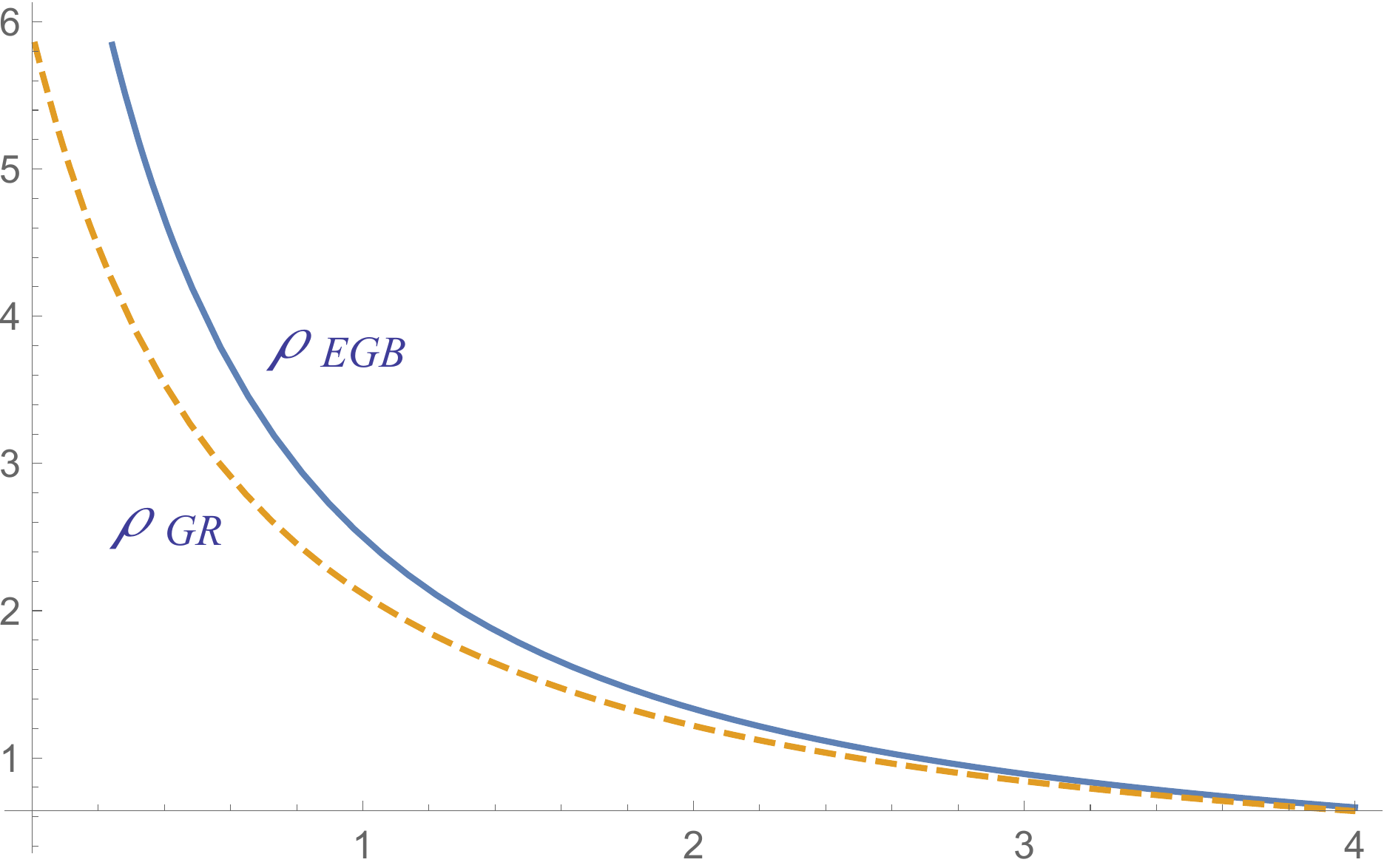}\\
  \caption{Plot of energy density $\rho/C$ versus radial value $x$}\label{Fig. 1}
\end{figure}

\begin{figure}
  \includegraphics[width=8cm]{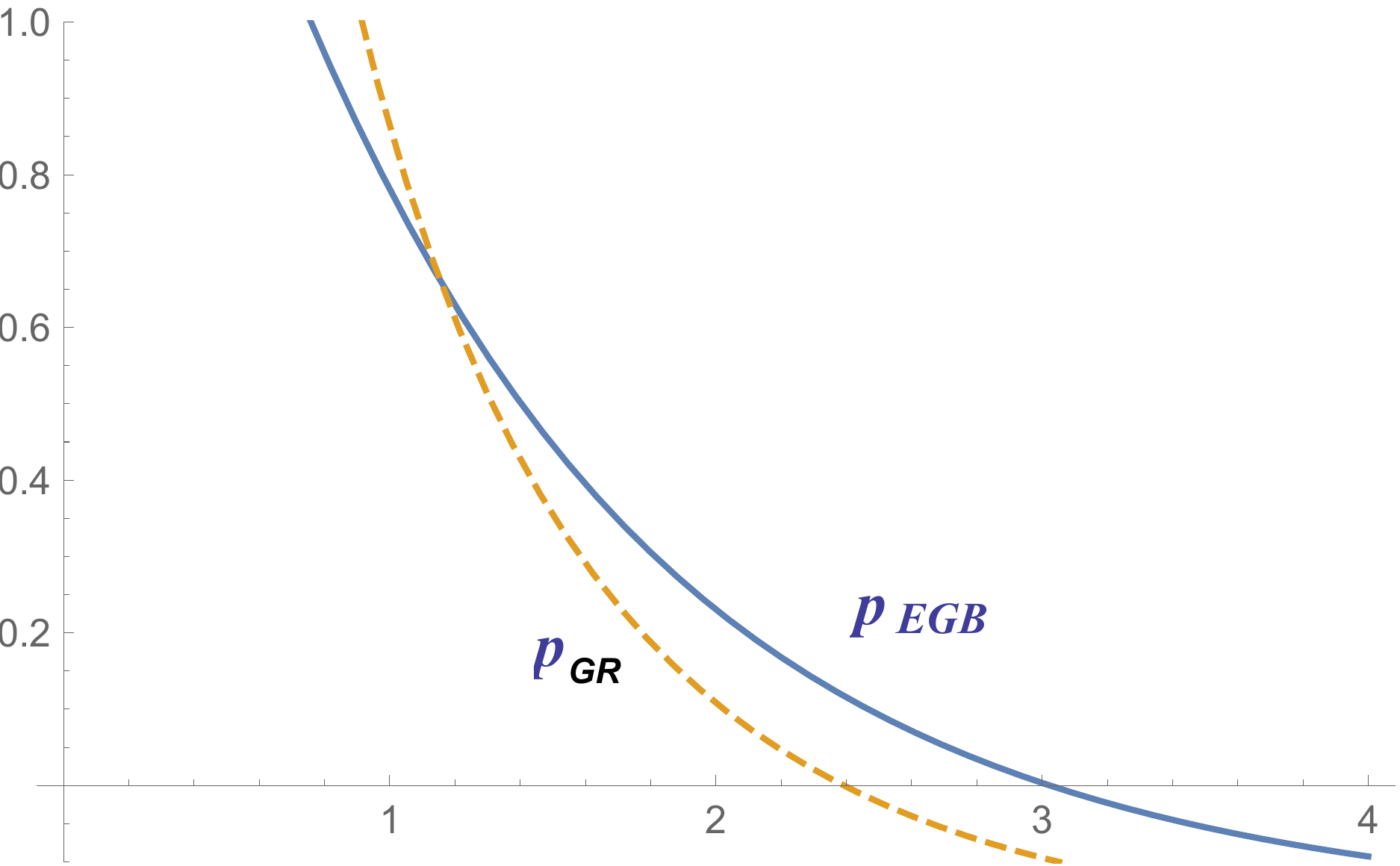}\\
  \caption{Plot of pressure $p/C$ versus radial value $x$}\label{Fig. 2}
\end{figure}

\begin{figure}
  \includegraphics[width=8cm]{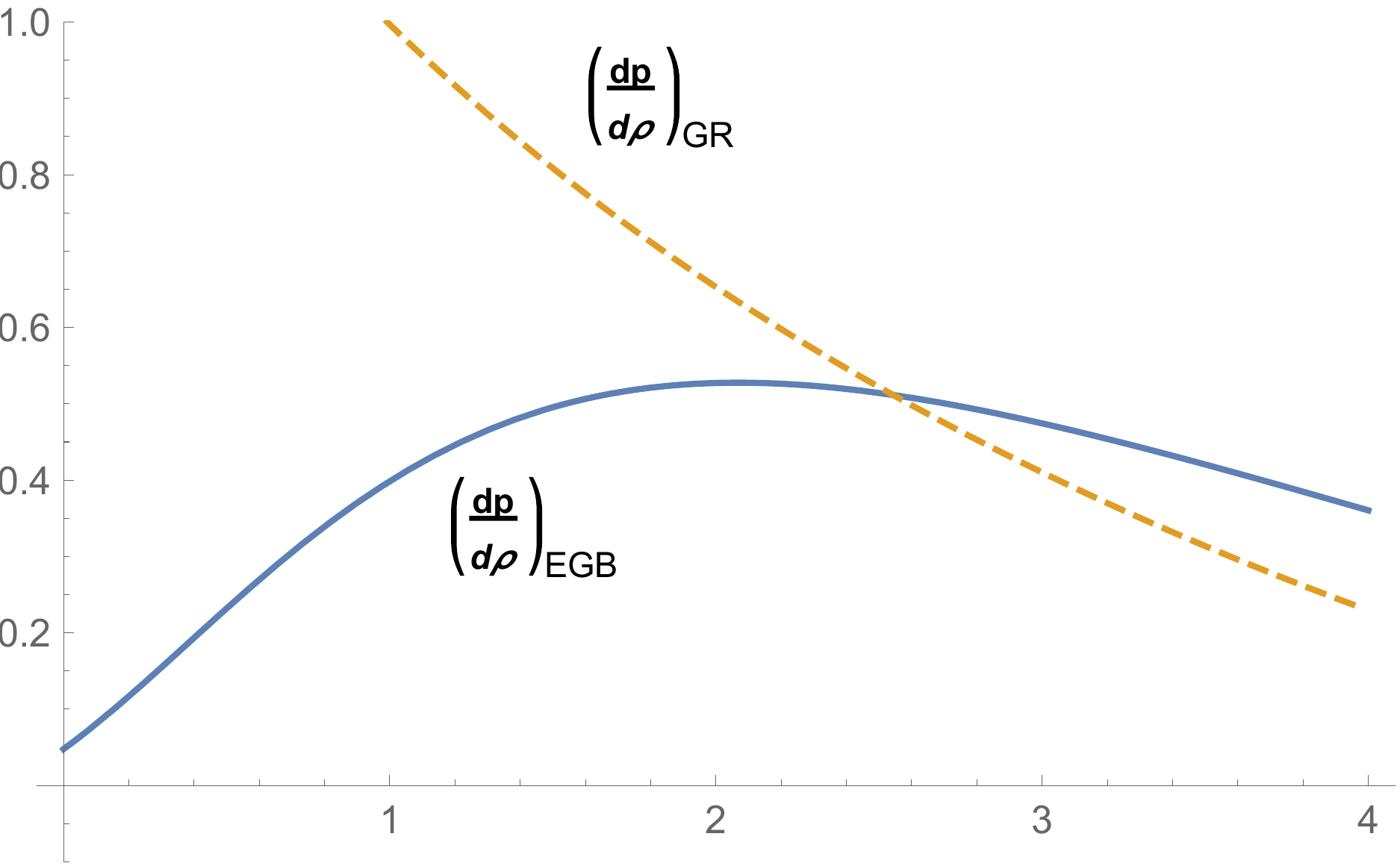}\\
  \caption{Plot of sound speed $\frac{dp}{d\rho}$ versus radial value $x$}\label{Fig. 3}
\end{figure}

\begin{figure}
  \includegraphics[width=8cm]{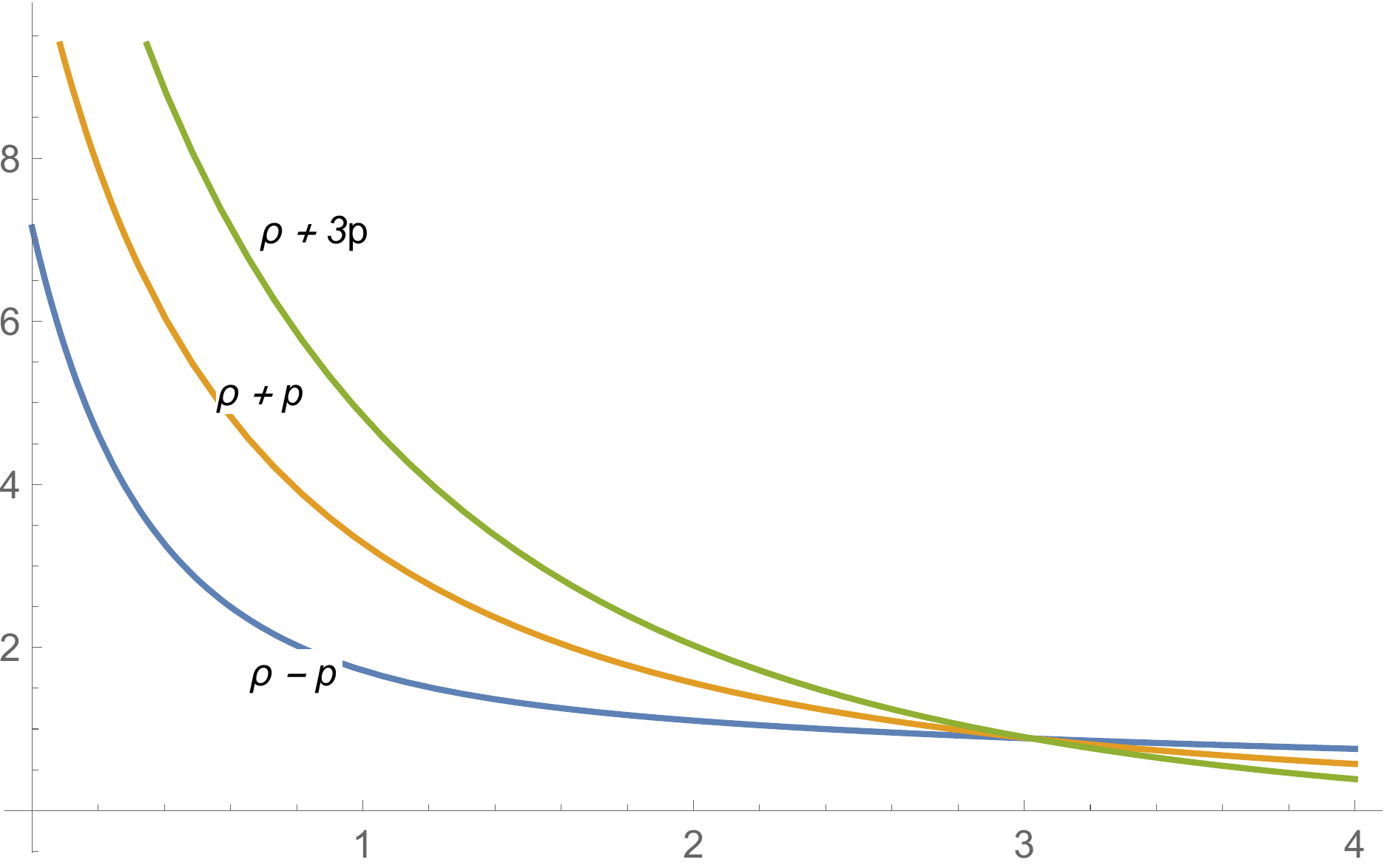}\\
  \caption{Plot of EGB energy conditions $\rho - p, \rho + p, \rho + 3p$ versus radial value $x$}\label{Fig. 4}
\end{figure}

\begin{figure}
  \includegraphics[width=8cm]{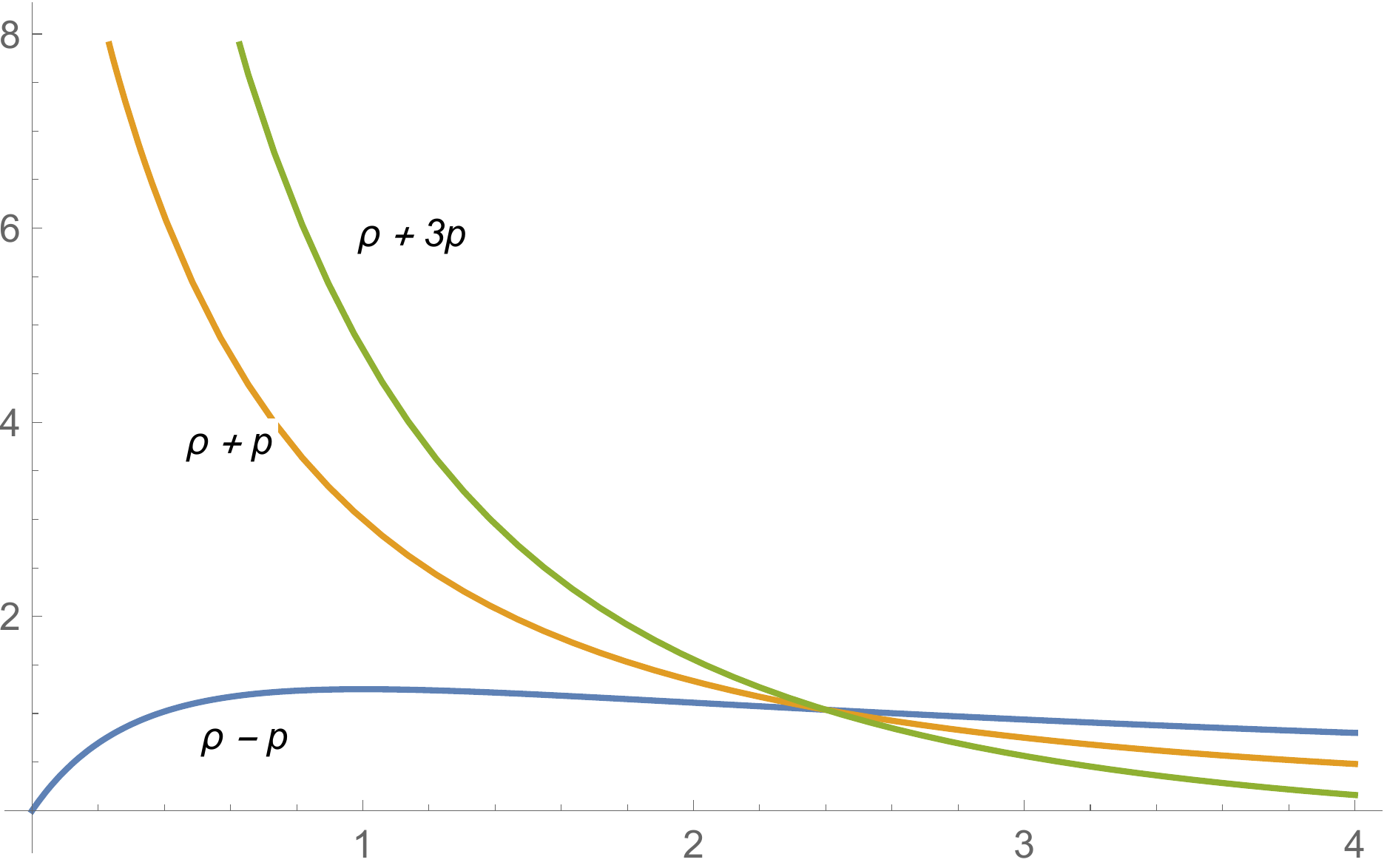}\\
  \caption{Plot of GR energy conditions $\rho - p, \rho + p, \rho + 3p$ versus radial value $x$}\label{Fig. 5}
\end{figure}

\begin{figure}
  \includegraphics[width=8cm]{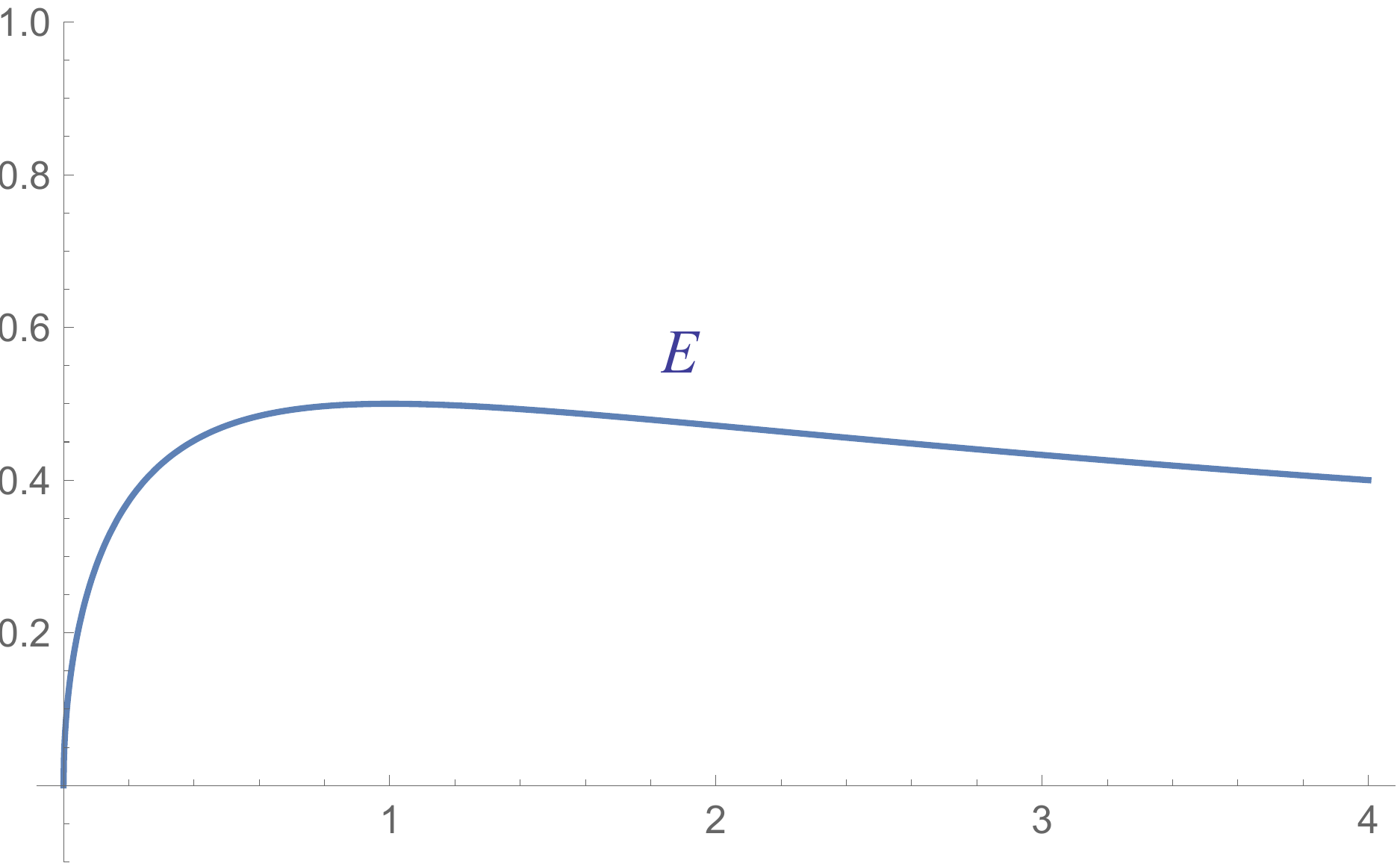}\\
  \caption{Plot of electric field intensity $E^2/C$ versus radial value $x$}\label{Fig. 6}
\end{figure}

\begin{figure}
  \includegraphics[width=8cm]{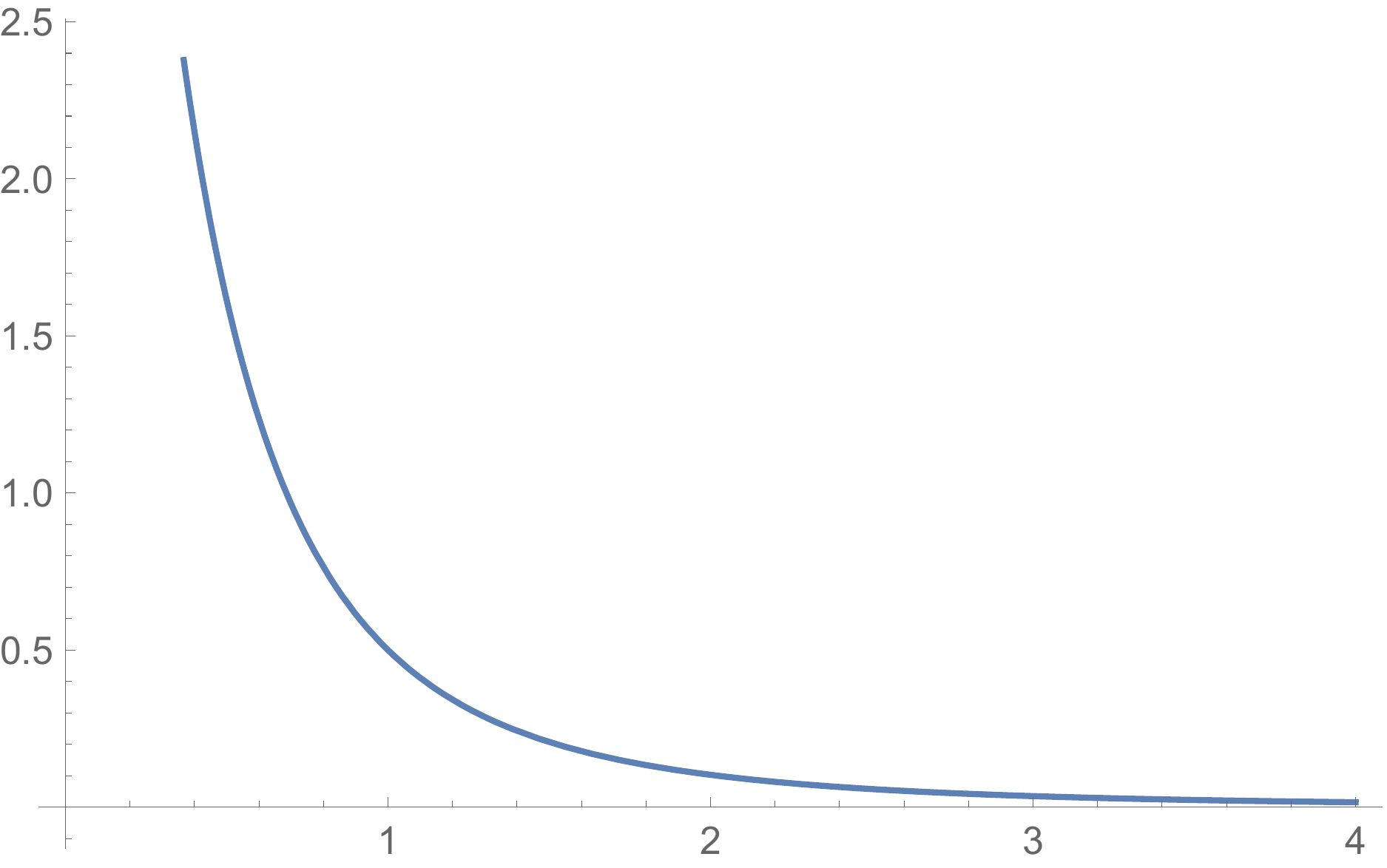}\\
  \caption{Plot of proper charge density $\sigma^2/C$ versus radial value $x$}\label{Fig. 7}
\end{figure}

\begin{figure}
  \includegraphics[width=8cm]{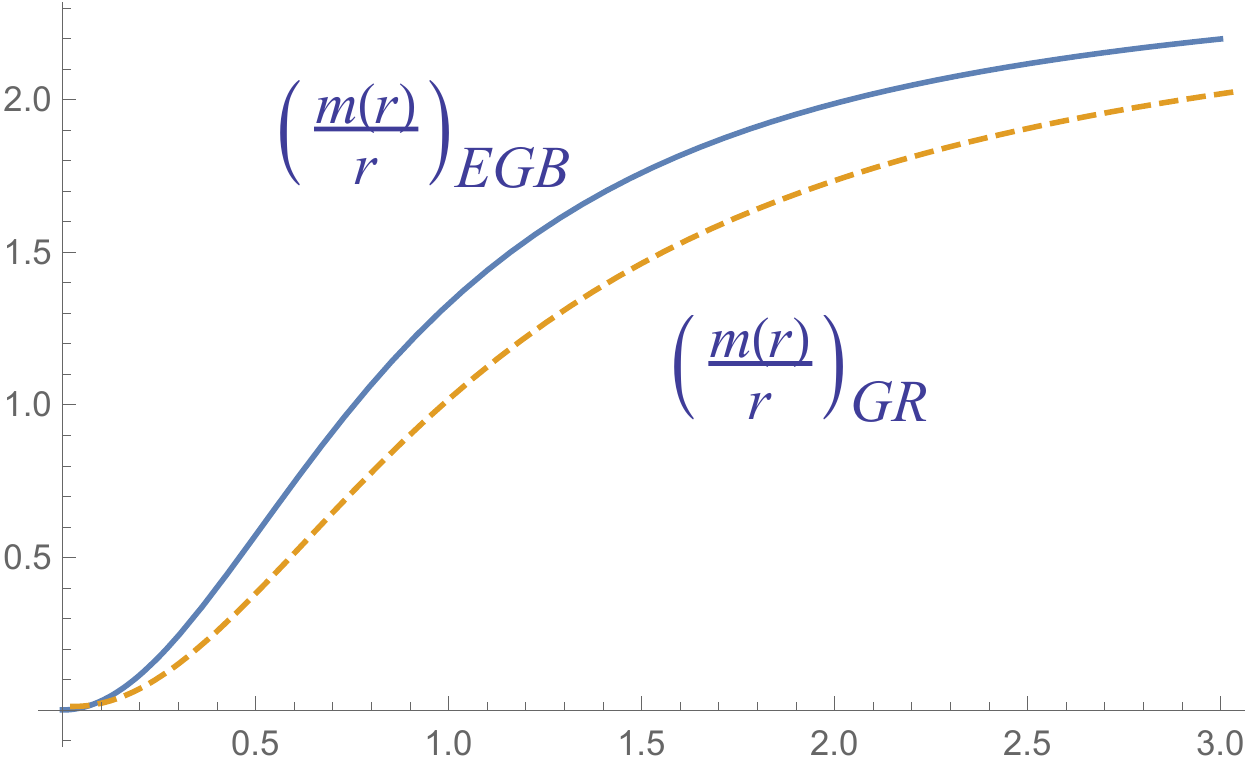}\\
  \caption{Plot of compactification function $\frac{m(r)}{r}$ versus radial value $r$}\label{Fig. 8}
\end{figure}

The plots display some important information about the viability of our model to represent realistic distributions of charged fluids. The pressure plot Fig. 2 reveals that the pressure is zero for the radial value $x=3$  units (EGB) and $x=2,5$ units (GR). This suggests that spheres of larger radius are admitted through the introduction of higher derivative terms in the lagrangian density. Additionally note that the pressure decreases monotonically outwards from the centre for both EGB and GR. Within these radii, Fig. 1 shows that the energy density is always positive and also has a negative gradient. Moreover for the same radius the sphere in EGB has a greater density in general.  Fig. 3 demonstrates that the sound speed remains lower than the light speed in that $0 < \frac{dp}{d\rho} < 1$ within the boundary for the EGB case. On the other hand, the graphs suggest that causality is violated in the radial interval $0 < x < 1$ for the GR framework. The energy conditions are plotted in Fig 4 and Fig 5 and we can observe that the weak, strong and dominant energy conditions are satisfied everywhere within the fluid for both gravity models: $\rho - p > 0, \rho + p > 0$ and $\rho + 3p > 0$. From Fig. 6 and Fig. 7 it can be noted that the electric field increases for a while and then slowly drops off towards the boundary and the proper charge density is always positive and a smoothly decreasing function outwardly. Since $E$ and $\sigma$ are independent of $y$, the plots for EGB and GR are the same. Finally Fig. 8 demonstrates a higher compactification values for the function $\frac{m(r)}{r}$ in the EGB framework when compared to GR. Interestingly both compactification functions have the same $\displaystyle{\lim_{r\rightarrow \infty}} = \frac{5}{2}$.  All of these are reasonable conditions for a perfect fluid sphere in an Einstein--Maxwell field. Clearly the higher curvature terms have a marked impact on the gravitational behaviour of the static charged fluid sphere and the indications are that undesirable features such as causality violation may be eliminated through the introduction of higher curvature terms.

\subsection{Schwarzschild metric ansatz}

Setting $a=1$ and $b=0$ generates the Schwarzschild potential $Z=1+x$. The solution (\ref{002}) simplifies to
\be
y=c_1\sqrt{1+x} + c_2 \label{006}
\ee
which is also identical to the remaining Schwarzschild potential for a constant density perfect fluid in Einstein gravity. The reason for this is that for this choice of $Z$ the electric field vanishes. This means that a different functional form for $E$ should be chosen to obtain a model of a charged fluid. This will be pursued in a different article.

\subsection{Isothermal type potential}

The case of a constant gravitational potential $Z = B$ for $B$ some constant, generates the solution
\be
y= c_1 + c_2\left(x+k(B-1)\ln (k(1-B)+x) \right) \label{005}
\ee
The complete solution for the energy density, pressure, charge density and speed of sound respectively is given by
\beq
\frac{\rho}{C} &=& \frac{5 (1-B)}{2 x} \label{0051a} \\ \n \\
\frac{p}{C} &=& \frac{5 (B-1)^2 c_1 k \log (k(1-B)+x)+(17 B-5) c_1 x+5 (B-1) c_2}{2 x \left((B-1) c_1 k \log (k(1-B)+x)+c_1 x+c_2\right)}  \label{0051b} \\ \n \\
\frac{\sigma^2}{C^2} &=& \frac{(1-B) B }{x^2}  \label{0051c} \\ \n \\
\frac{E^2}{C} &=& \frac{1-B}{x} \label{0051d} \\ \n \\
\frac{dp}{d\rho} &=& \frac{12 B c_1^2 x^3}{5 (B-1) ((B-1) k-x) \left((B-1) c_1 k \log (k(1-B)+x)+c_1 x+c_2\right){}^2}-1 \label{0051e}
\eeq
Demanding a positive energy density requires $B<1$ from (\ref{0051a}) while ensuring the right hand side of (\ref{0051c}) and (\ref{0051d}) remains positive restricts $B$ as $0 < B < 1$. Therefore this model will only be physically reasonable if $0<B<1$.  The vanishing of the pressure is also possible although this requires solving a non-algebraic equation - however, such a radial value exists demarcating the boundary of the fluid. Observe that the energy density obeys the inverse square law $\rho \sim \frac{1}{r^2}$ which is characteristic of isothermal fluid spheres in Newtonian gravity. However, the linear barotropic  equation of state ($p \sim \rho$) is not in effect here. An equation of state does indeed exist for this solution as $x$ may easily be found in terms of $\rho$ via (\ref{0051a}) and this may then be substituted in (\ref{0051b}) to give the equation of state $p=p(\rho)$ explicitly.  It has also been established by Dadhich {\it et al} \cite{isothermal}  that a constant gravitational potential is a necessary and sufficient condition for isothermal behaviour for pure Lovelock gravity where the action is constructed using the $N^{th}$ order term in the Lovelock polynomial.  To ensure a subluminal sound speed, it is required that (\ref{0051e}) lie between 0 and 1. The forms of the expressions do not allow for an analytic treatment and the use of graphical representations may prove useful. The vanishing of the surface pressure at the boundary $r = R$ together with the matching of gravitational potentials $g_{00}$ of the charged Boulware--Deser (\ref{008})  metric fix the values of the integration constants as
\beq
c_1 &=&  -\frac{5 (B-1) V}{12 \alpha  \left((5-17 B) C R^2+5 (B-1) \text{CR}^2\right)}  \\ \n \\
c_2 &=& \frac{\left(5 (B-1)^2 k \log \left(-B k+C R^2+k\right)+(17 B-5) C R^2\right) V}{12 \alpha  \left((5-17 B) C R^2+5 (B-1) \text{CR}^2\right)}
\eeq
in terms of the mass $M$ and charge $Q$ of the fluid and where we have put \\  $V=R^2 \left(\sqrt{\frac{72 \alpha  M}{R^4}-\frac{24 \alpha  Q^2}{R^6}+9}-3\right)-12 \alpha K$ for simplicity.

\section{Discussion}

We have written the EMGB field equations governing the behaviour of a static ball of perfect fluid matter in the presence of an electric field. Exact solutions of the field equations were obtained by prescribing one of the metric potentials as well as by prescribing the electric field intensity to behave dimensionally as the reciprocal of the radius. The metric ansatz utilised generalised the spheroidal geometry discussed by Vaidya and Tikekar  for their superdense star models. Special cases included the Vaidya--Tikekar \cite{vt}  geometry, the Finch--Skea \cite{fs} metric, the Schwarzschild interior metric as well as the constant gravitational potential fluid sphere. The Schwarzschild choice coupled with the electric field prescribed failed to produce a charged distribution. A different electric field intensity must be selected to construct an EMGB model. In the remaining cases, it was possible to explicitly obtain exact solutions yielding the metric potentials, the energy density, pressure, electric field intensity and proper charge density. It was also demonstrated that in each case a hypersurface of zero pressure existed identifying the fluids boundary. Across this boundary the matching of gravitational potentials allowed to the settling of all integration constants. In the case of the Finch--Skea ansatz it was demonstrated graphically that suitable parameter values existed  in order to generate a model that satisfied elementary physical requirements. In particular, it was shown that the sound speed always remained lower than the speed of light so that causality was maintained. The higher curvature terms support spheres of greater radius and density in the EGB regime when compared to the 5-dimensional Einstein case. Moreover, the GB term ensured that the model was causal while the Einstein case suffered the defect of the fluid being superluminal from the centre to a radial value well inside the boundary. This illustrates the potential of the Gauss--Bonnet higher curvature terms in correcting the physical behaviour of realistic objects which is a clear improvement on general relativity.  Importantly, it has been shown in each case that a barotropic equation of state exists in the EGB framework.  This investigation therefore leads us to conclude that compact star models indeed do exist in the EMGB framework and that GB higher curvature terms improve the likelihood of models conforming to realistic distributions. The important open questions under investigation include the behaviour of simple charged dust as well as the consequences of imposing a linear equation of state at the outset on the nonlinear system. In addition it will be useful to check the impact of the higher curvature terms on the properties of compact objects such as neutron stars and fluid planets. These studies are presently being undertaken.




\begin{thebibliography}{}
\bibitem{hm} S Hansraj, B Chilambwe and S D Maharaj, {\em Eur. Phys. J. C } {\bf 27} 277 (2015)
\bibitem{mh} S D Maharaj, B. Chilambwe and S Hansraj,  {\em Phys. Rev. D} {\bf 91},  084049 (2015)
\bibitem{chil-hans} B Chilambwe, S. Hansraj and S D Maharaj, {\em Int. J. Mod. Phys. D} {\bf 24} 1550051 (2015)
\bibitem{boulware} D G Boulware and S Deser,{\em Phys. Rev. Lett.} {\bf 55} 2656 (1985)
\bibitem{wilt} D L Wiltshire, {\em Phys. Rev. D} {38}, 2445 (1988)
\bibitem{ghosh} S. G. Ghosh and S. Jhingan {\em Phys. Rev. D} {\bf  82}, 024017 (2010)
\bibitem{jhingan} S. Jinghan and S. G. Ghosh {\em Phys.Rev. D} {\bf 81}  024010 (2010) arXiv:1002.3245
\bibitem{blackstring} B Kleihaus, J Kunz and E Radu {\em Phys. Lett. B} {\bf 713} 110 (2012)
\bibitem{anabalon} A. Anabalón, N. Deruelle, Y. Morisawa, J. Oliva, M Sasaki, D. Tempo and R. Troncoso {\em Class. Quant. Grav.} {\bf 26} 065002 (2009)
\bibitem{dad-hans} N. Dadhich, S. Hansraj and B Chilambwe arXiv:1607.07095
\bibitem{seager} S Seager, M Kuchner, C A Hier--Majumder and B Militzer {\em The Astrophysical Journal}{\bf 669} 1279 (2007)
\bibitem{sato} B Sato, {\it {et al.}}, {\em ApJ}, {\bf 633}, 465 (2005)
\bibitem{butler} R P Butler, S S Vogt, G W  Marcy, D A  Fischer, J T Wright, G W Henry, G Laughlin, and J J Lissauer, {\em  ApJ}, {\bf 617}, 580 (2004)
\bibitem{gillon} M Gillon, F Pont, B O Demory, F Mallmann, M Mayor, T Mazeh, D Queloz, A Shporer, S Udry and C. Vuissoz {\em Astronomy and Astrophysics} {\bf 472} L13 (2007)
\bibitem{ivan} B V Ivanov, {\it Phys. Rev. D } {\bf 65}  104001 (2002)
\bibitem{lovelock} D Lovelock, {\em J. Math. Phys.} {\bf 12} 498 (1971).
\bibitem{whit} B Whitt, {\em Phys. Rev. D} {\bf 38}, 3000 (1988).
\bibitem{whee} J T Wheeler, {\em Nucl. Phys. B} {\bf 268}, 737 (1986); {\bf 27}, 732 (1986)
\bibitem{Ban1} M Ba\~{n}ados, C Teitelboim and J Zanelli, {\em Phys. Rev. D}, {\bf 49} 975 (1994)
\bibitem{Ban2} M Ba\~{n}ados, C Teitelboim and J Zanelli, {\em Phys. Rev. Lett.}, {\bf 69} 1849 (1992).
\bibitem{Myers} R C Myers and J Simon, {\em Phys. Rev. D} {\bf 38}, 2434 (1988)
\bibitem{probes3} N Dadhich, K Prabhu, J Pons,  {\em Gen. Relativ. Gravit.} {\bf 45} 1131 (2013)
\bibitem{cai} R G Cai and N Ohta, {\em  Phys. Rev. D} {\bf 74}  064001 (2006)
\bibitem{Sharm} J P Sharma, {\em Astophysics and Space Science}, {\bf 163} 109 (1990)
\bibitem{dkm} N Dadhich, A Molina, A Khugaev, {\em Phys. Rev. D} {\bf 81}, 104026 (2010); arxiv:1001.3952
\bibitem{dad} N. Dadhich {\em Eur. Phys. J. C} {\bf 76} 104 (2016)
\bibitem{ellis1} G F R Ellis, H van Elst, J. Murugan and J-P Uzan {\em  Class. Quantum Grav.} {\bf 28} 225007 (2011)
\bibitem{ellis2} G F R Ellis, {\em Gen. Relativ. Gravit.}  {\bf 46} 1619 (2014)
\bibitem{weinberg} S Weinberg {\em  Rev. Mod. Phys.} {\bf 61} 1 (1989)
\bibitem{staro}  A. A. Starobinsky, {\em Phys. Lett. B} {\bf 91} 99 (1980)
\bibitem{harko}  Harko, T., Lobo, F.S.N., Nojiri, S. and Odintsov, S.D. {\em Phys. Rev. D} {\bf 84} 024020 (2011)
\bibitem{krori} K D Krori and J Barua {\em J. Phys. A: Mathematical and General} {\bf 8} 508 (1975)
\bibitem{hansraj} S Hansraj and S D Maharaj {\em Int. J. Mod. Phys. D} {\bf 15} 1311 (2006)
\bibitem{gross} D Gross {\em Nucl. Phys. Proc. Suppl.} {\bf  74} 426 (1999)
\bibitem{herrera} L. Herrera  and J. J. Ponce de Leon {\em J. Math. Phys.}, {\bf 26}  2302 (1985)
\bibitem{davis} S C Davis, {\em Phys. Rev. D} {\bf 67}, 024030 (2002)
\bibitem{nils1} Nilsson U S and Uggla C  {\em Annals of Physics} {\bf 286}  278 (2000)
\bibitem{nils2} Nilsson U S and Uggla C  {\em Annals of Physics} {\bf 286}  292 (2000)
\bibitem{vt} PC Vaidya and R Tikekar, {\em J. Astrophys.} {\bf}3, 325 (1982)
\bibitem{fs} M R Finch and J E F Skea {\em Class. Quant. Grav.} {\bf 6} 467 (1989)
\bibitem{isothermal} N. Dadhich, S Hansraj, S D Maharaj, {\em Phys. Rev. D} {\bf 93}  044072 (2016)
\bibitem{sas} W. C. Saslaw, S. D. Maharaj and N. Dadhich {\em The Astrophys. J.} {\bf 471} 571 (1996)
\bibitem{wal} J D Walecka ,  {\em Phys. Lett. B} {\bf 59} (1975) 109
\end{thebibliography}

\end{document}